\documentstyle[aps,epsf]{revtex}
\draft

\begin{document}

\title{Looking at friction through ``shearons''}

\author{Markus~Porto, Michael~Urbakh, and Joseph~Klafter}

\address{School~of~Chemistry, Tel~Aviv~University, 69978~Tel~Aviv, Israel}

\date{July 6, 1999}

\maketitle

\begin{abstract}
We study the response to shear of a one-dimensional monolayer embedded between
two rigid plates, where the upper one is externally driven. The shear is shown
to excite ``shearons'', which are collective modes of the embedded system with
well defined spatial and temporal pattern, and which dominate the frictional
properties of the driven system. We demonstrate that static friction,
stick-slip events, and memory effects are well described in terms of the
creation and/or annihilation of a shearon. This raises the possibility to
control friction by modifying the excited shearon, which we examplify by
introducing a defect at the bottom plate.
\end{abstract}

\bigskip

The field of nanotribology evolves around the attempts to understand the
relationship between macroscopical frictional response and microscopic
properties of sheared systems \cite{Bowden/Tabor:1985+Persson:1998}. New
experimental tools such as the surface force apparatus (SFA) are used to
explore shear forces between atomically smooth solid surfaces separated by a
nanoscale molecular film
\cite{Yoshizawa/McGuiggan/Israelachvili:1993+Hu/Carson/Granick:1991+%
Klein/Kumacheva:1995+Berman/Drucker/Israelachvili:1996}. These experiments have
unraveled a broad range of phenomena and new behaviors which help shed light on
some ``old'' concepts which have been considered already textbook materials:
static and kinetic friction forces, transition to sliding, thinning, and memory
effects. These and other observations have motivated theoretical efforts
\cite{Tomlinson:1929,Tompson/Robbins/Grest:1995,Rozman/Urbakh/Klafter:1996,%
Carlson/Batista:1996+Batista/Carlson:1998,Braiman/Family/Hentschel:1996-1997,%
Gao/Luedtke/Landman:1997,%
Rozman/Urbakh/Klafter:1997+Rozman/Urbakh/Klafter/Elmer:1998+%
Zaloj/Urbakh/Klafter:1998,Elmer:1998,Baumberger/Caroli:1998,%
Sokoloff:1995+Sokoloff/Tomassone:1998}, both numerical and analytical, but many
issues have remained unresolved, in particular the relation between the
macroscopic observables and the microscopic properties of the embedded system.

In this Letter we introduce the concept of ``shearons'', which are
shear-induced collective modes excited in the embedded system and characterized
by their wave vector $q$. Shearons, which display well defined spatial and
temporal patterns, dominate the frictional properties of the driven system and
are found useful in establishing a connection between frictional response and
motional modes of the embedded system. Within this framework observations such
as static friction, stick-slip behavior, and memory effects can be correlated
with the creation and/or annihilation of a shearon. These correlations suggests
the possibility to control friction by modifying the shearon's wave vector and
thereby tuning the embedded system, by adding, for example, a defect at one of
the plates.

We start from a {\it microscopic} model which has been investigated recently
and has been shown to capture many of the important experimental findings
\cite{Rozman/Urbakh/Klafter:1996,%
Rozman/Urbakh/Klafter:1997+Rozman/Urbakh/Klafter/Elmer:1998+%
Zaloj/Urbakh/Klafter:1998}. The model system consists of two rigid plates, with
a monolayer of $N$ particles with masses $m$ and coordinates $x_i$ embedded
between them. The top plate with mass $M$ and center of mass coordinate $X$ is
pulled with a linear spring of spring constant $K$. The spring is connected to
a stage which moves with velocity $V$. This system is described by $N+1$
equations of motion
\begin{equation}\label{eq:topplate}
M \ddot{X} +
\sum_{i = 1}^N \eta (\dot{X} - \dot{x}_i) +
K (X-V t) +
\sum_{i = 1}^N \frac{\partial \Phi(x_i - X)}{\partial X} = 0
\end{equation}
\begin{equation}\label{eq:particles}
m \ddot{x}_i +
\eta (2 \dot{x}_i - \dot{X}) +
\sum_{{\scriptstyle j = 1} \atop {\scriptstyle j \not= i}}^N
\frac{\partial \Psi(x_i-x_j)}{\partial x_j} +
\frac{\partial \Phi(x_i)}{\partial x_i} +
\frac{\partial \Phi(x_i-X)}{\partial x_i} = 0 \quad i = 1, \ldots, N.
\end{equation}
The second term in Eqs.~(\ref{eq:topplate}) and (\ref{eq:particles}) describes
the dissipative forces between the particles and the plates and is proportional
to their relative velocities with proportionality constant $\eta$, accounting
for dissipation that arises from interactions with phonons and/or other
excitations. The interaction between the particles and the plates is
represented by the periodic potential $\Phi(x) = -\Phi_0 \cos(2 \pi x/b)$.
Concerning the inter-particle interaction, we assume nearest neighbor
interactions of two types: {\it (i)}~harmonic interaction $\Psi(x_i - x_{i \pm
1}) = (k/2) [x_i - x_{i \pm 1} \pm a]^2$ and {\it (ii)}~Lennard-Jones
interaction $\Psi(x_i - x_{i \pm 1}) = (k a^2/72) \{[a/(x_i - x_{i \pm
1})]^{12}-2 [a/(x_i - x_{i \pm 1})]^6\}$ \cite{Note1}. The two plates do not
interact directly.

The basic frequency is chosen to be the frequency of the top plate oscillation
in the potential $\Omega \equiv (2 \pi/b) \sqrt{N \Phi_0/M}$. The other
frequencies in the model are the frequency of the particle oscillation in the
potential $\omega \equiv (2 \pi/b) \sqrt{\Phi_0/m}$, the characteristic
frequency of the inter-particle interaction $\hat{\omega} \equiv \sqrt{k/m}$,
and the frequency of the free oscillation of the top plate $\hat{\Omega} \equiv
\sqrt{K/M}$. To simplify the discussion we introduce unitless coordinates $Y
\equiv X/b$ and $y_i \equiv x_i/b$ of the top plate and the particles,
respectively, as well as an unitless time $\tau \equiv \Omega t$. We define the
following quantities: The misfit between the substrate and inter-particle
potentials' periods $\Delta \equiv 1 - a/b$, the ratio of masses of the
particles and the top plate $\epsilon \equiv N m/M$, the unitless dissipation
coefficient $\gamma \equiv N \eta/(M \Omega)$, the ratio of frequencies of free
oscillation of the top plate and the oscillation of the top plate in the
potential $\alpha \equiv \hat{\Omega}/\Omega$, the ratio of the frequencies
related to the inter-particle and particle/plate interactions $\beta \equiv
\hat{\omega}/\omega$, and the dimensionless velocity $v \equiv V/(b \Omega)$.
Additionally, the friction force per particle $f_{\rm k}$ is defined as $f_{\rm
k} \equiv F_{\rm k}/(N F_{\rm s})$, where $F_{\rm k}$ is the total friction
force measured with the external spring and $F_{\rm s} \equiv 2 \pi \Phi_0/b$
is the force unit given by the plate potential. Here, we keep fixed the number
of particles $N = 15$, the mass ratio $\epsilon = 0.01$, the misfit $\Delta =
0.1$, and the relative strength of the inter-particle interaction $\beta^2 =
1$. We vary only the relative strength of the external spring $\alpha^2$, the
dissipation coefficient $\gamma$, and stage velocity $v$.

In order to analyze the motion of the embedded system more closely, we seperate
the motion of the particles into the center of mass part $y_{\rm cms} \equiv
1/N \sum_{i = 1}^N y_i$ \cite{Note2} and the fluctuations $\delta y_i \equiv
y_i - y_{\rm cms}$. It has been observed in similar models
\cite{Smith/Robbins/Cieplak:1996,Braiman/Hentschel/Family/Mak/Krim:1999,%
Hentschel/Family/Braiman:1999} that different modes of motion can coexist for a
given set of parameters and lead to different frictional forces. Here, we
concentrate on these solutions of the coupled dynamical
equations~(\ref{eq:topplate}) and (\ref{eq:particles}) which correspond to
smooth or to stick-slip motion of the top plate. To understand the nature of
the motion of the embedded system in these regimes and the relation to the
frictional response, we choose the particles density $\varrho$ as a
observable. Instead of defining the density as $\sum_{i = 1}^N \delta(y-y_{\rm
cms}-\delta y_i)$, we represent each particle by a Gaussian of width $\sigma =
1$, namely
\begin{equation}\label{eq:density}
\varrho(y-y_{\rm cms}, \tau) \equiv \sum_{i = 1}^N \exp
\left\{ -\left[ \frac{y-y_{\rm cms}-\delta y_i}{\sigma} \right]^2 \right\}
\quad.
\end{equation}
This allows to visualize correlated motions of the particles on length scales
which exceed nearest neighbor distance and are of the order of $2 \sigma$ to $3
\sigma$. Using Eq.~(\ref{eq:density}) we find that the response of the embedded
system to shear can be described in terms of collective modes which result in
well defined spatial and temporal pattern in the density, which we call
``shearons''. These shearons are the spatial/temporal manifestation of
parametric resonance between the external drive and the embedded system
\cite{Hentschel/Family/Braiman:1999,Weiss/Elmer:1997}.

\begin{figure}[h]
\nopagebreak\vspace*{5mm}\nopagebreak
\def\epsfsize#1#2{0.619#1}
\noindent\hfill\epsfbox{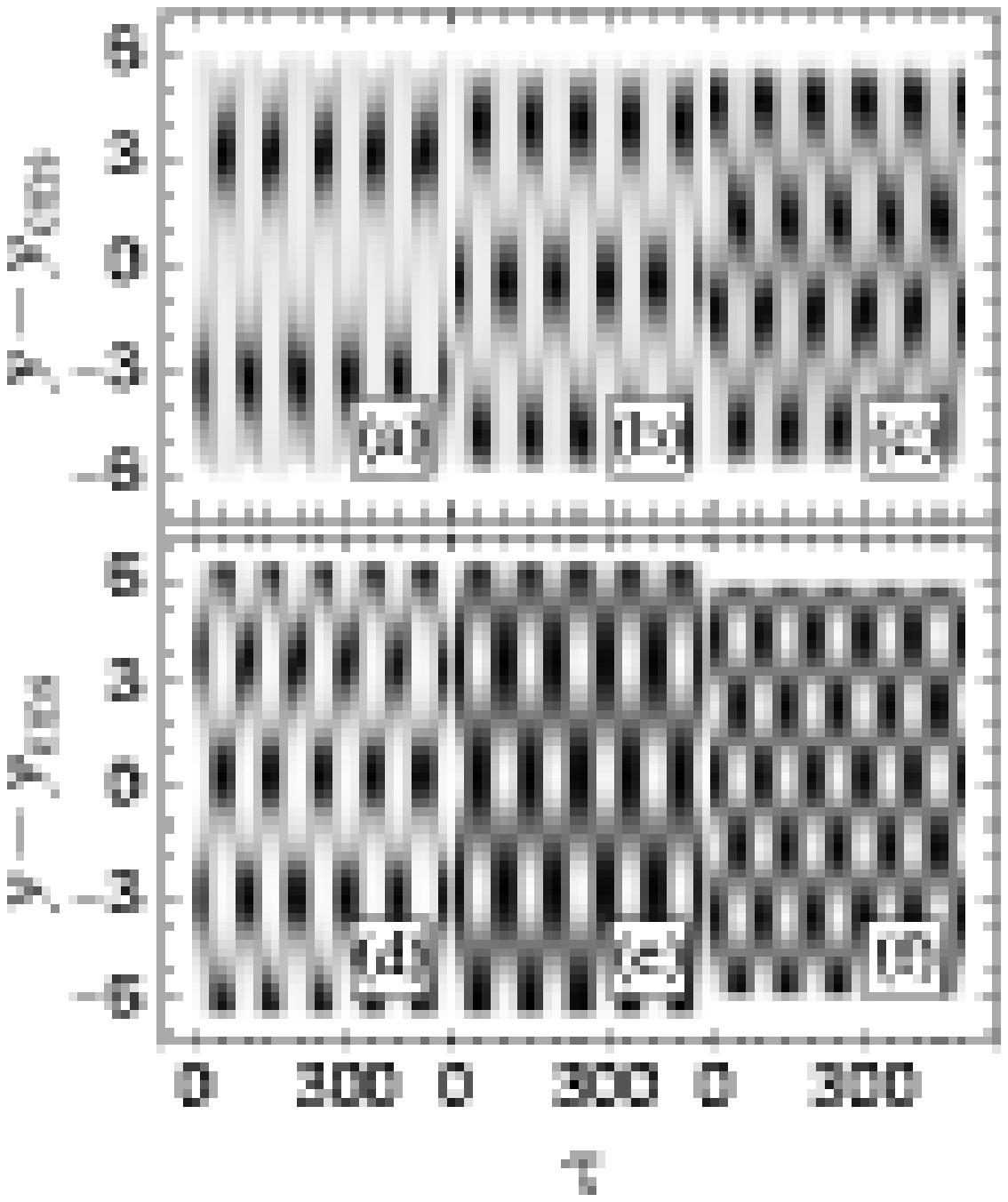}\hfill\hspace*{0mm}%
\nopagebreak\vspace*{5mm}\nopagebreak
\caption{Plot of the particles density $\varrho$ vs position $y-y_{\rm
cms}$ and time $\tau$ for different shearons. The white and black colors
indicate low and high density, respectively [the gray scales are chosen
independently for each subfigure to maximise contrast]. In (a)-(c) three
different stable shearons are shown for a chain with free ends. In
(d)-(f) periodic boundary conditions are used with box sizes $\ell = 13$,
$13$, and $12$. Harmonic interaction between the particle is assumed in
(a)-(d) and Lennard-Jones interaction in (e) and (f). The resulting mean
friction forces in (a)-(f) are $\left< f_{\rm k} \right> \protect\cong
0.263$, $0.180$, $0.125$, $0.123$, $0.060$, and $0.047$. The model
parameters are $\alpha = 1$, $\beta = 1$, $\gamma = 0.75$, $\Delta
= 0.1$, $\epsilon = 0.01$, $N = 15$, and $v = 0.02$. [Please note that
the figure has a reduced quality due to figure size limitations of the
condmat database.]}
\label{figure:shearons}
\end{figure}

In Fig.~\ref{figure:shearons}(a)-(c) we present three {\it stable} shearons for
a chain of harmonically interacting particles with free boundary conditions,
where in all cases the top plate slides smoothly. All three shearons share the
same set of parameters and differ only in their initial conditions. We find
that the observed mean friction force per particle $\left< f_{\rm k} \right>$
($\left< \cdot \right>$ denotes time average), which is directly related to the
spatial/temporal fluctuations by \cite{Weiss/Elmer:1997}
\begin{equation}\label{eq:friction}
\left< f_{\rm k} \right> = \pi \gamma v + \frac{2 \pi \gamma}{v}
\left< \sum_{i=1}^N \delta\dot{y}_i^2 \right>
\quad,
\end{equation}
decreases with increasing shearon wave vector $q$. This means that the mean
friction force can be reduced by increasing the shearon wave vector. In the
present example, we observe a reduction from $\left< f_{\rm k} \right> \cong
0.263$ [Fig.\ref{figure:shearons}(a)] to $\left< f_{\rm k} \right> \cong 0.180$
[Fig.\ref{figure:shearons}(b)], and to $\left< f_{\rm k} \right> \cong 0.125$
[Fig.\ref{figure:shearons}(c)]. The lowest possible mean friction force $\left<
f_{\rm k} \right> = \pi \gamma v$ can be achieved in the limit $q \to \infty$.
For our choice of parameters $\pi \gamma v \cong 0.04712$.

We find that the regions of high density fluctuations exhibit low velocity
fluctuations and, according to Eq.~(\ref{eq:friction}), low dissipation. These
regions are surrounded and stabilized by regions of low density fluctuations
with highly fluctuating velocities and hence high dissipation. We note that the
minima/maxima of the density reflect correlations in the motion of the
neighboring and next-neighboring particles rather than actual locations. The
cooperative nature of motion is visualized here introducing a finite width
$\sigma$.

As we have already seen, larger wave vectors correspond to higher average
density of the embedded system. This means that for a given number of particles
the mean chain length $\left< L \right> \equiv \left< y_N-y_1 \right>$
decreases with $q$: $\left< L \right> \cong 13$
[Fig.~\ref{figure:shearons}(a)], $\left< L \right> \cong 12.5$
[Fig.~\ref{figure:shearons}(b)], and $\left< L \right> \cong 12$
[Fig.~\ref{figure:shearons}(c)]. Note that for $N = 15$ and $\Delta = 0.1$ the
equilibrium length of the free chain is $12.6$. The interplay between mean
length $\left< L \right>$ and the shearon wave vector $q$ becomes clearer
within the calculations under periodic boundary conditions, where $q$ is a
function of the box size $\ell$, see Fig.~\ref{figure:shearons}(d)-(f). In
Fig.~\ref{figure:shearons}(d) we present the spatial/temporal pattern found for
harmonically interacting particles within a box of size $\ell = 13$. It is
found that for similar wave vectors the mean friction force is smaller for the
case of periodic boundary conditions compared to a free chain [c.f.\
Fig.~\ref{figure:shearons}(c)], since the free ends strongly fluctuate and
cause additional dissipation.

When replacing the harmonic with the Lennard-Jones interaction [see
Fig.~\ref{figure:shearons}(e) and (f)], the picture remains basically
unchanged. The essential differences are found for: {\it (i)}~very large box
sizes, when the embedded Lennard-Jones system breaks and one reobtaines the
independend particle scenario described in \cite{Rozman/Urbakh/Klafter:1996};
{\it (ii)}~very small box sizes, when the hardcore of Lennard-Jones potential
becomes dominant. The latter case corresponds to the high pressure limit $p
\propto \ell^{-1} \to \infty$ and results in a strong stiffness. This, together
with incommensurability, allows to achive the lower bound for the mean friction
force $\left< f_{\rm k} \right> = \pi \gamma v$, c.f.\ Eq.~(\ref{eq:friction}).
In the case of harmonically interacting particles a minimum of the mean
friction force $\left< f_{\rm k} \right>$ as a function of box size $\ell$ is
found for a finite value of $\ell$, which is determined by commensurate
condition of shearon wave vector and the box size. These effects will be
discussed elsewhere \cite{Porto/Urbakh/Klafter:1999}.

\begin{figure}[h]
\nopagebreak\vspace*{5mm}\nopagebreak
\def\epsfsize#1#2{0.619#1}
\noindent\hfill\epsfbox{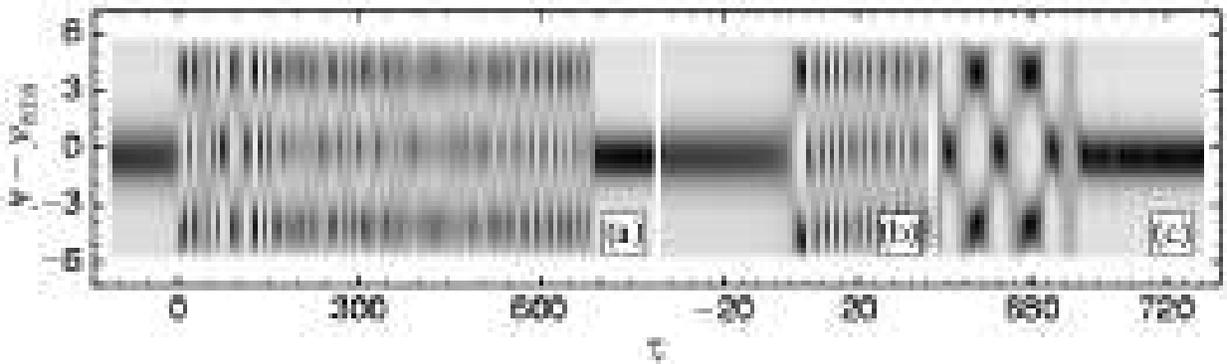}\hfill\hspace*{0mm}%
\nopagebreak\vspace*{5mm}\nopagebreak
\caption{Plot of the particles density $\varrho$ vs position $y-y_{\rm
cms}$ and time $\tau$ for parameters that correspond to periodic
stick-slip motion. In (a) we focus on the time interval of a single slip
event. The slip starts at $\tau \protect\cong 0$ [largest spring force]
and persits until $\tau \protect\cong 689$ [smallest spring force], (b)
and (c) show enlargements of the region where the motion of the top plate
starts and stops. The model parameters are $\alpha = 0.02$, $\beta = 1$,
$\gamma = 0.1$, $\Delta = 0.1$, $\epsilon = 0.01$, $N = 15$ [harmonic
interaction and free boundary conditions], and $v = 0.06$. [Please note that
the figure has a reduced quality due to figure size limitations of the
condmat database.]}
\label{figure:stick+slip}
\end{figure}

We now explore the concept of shearons in relationship to varios frictional
phenomenon. We start from the well known stick-slip phenomenon observed in many
nanoscale systems at low driving velocity, but whose nature is still not well
understood. The start of a slip event has been commonly attributed to the
`melting' of the embedded system, namely a transition from a ordered
`solid'-like to a disordered `liquid'-like structure, that `refreezes' at the
end of the slip event. In Fig.~\ref{figure:stick+slip}(a) we show that during
slippage the motion of the embedded system is {\it highly ordered} and {\it
highly correlated}. At the start of the slip event, the moment of the highest
spring force [enlarged in Fig.~\ref{figure:stick+slip}(b)], a shearon is
created, persisting with a constant wave vector until it gets annihilated at
the moment of the lowest spring force [enlarged in
Fig.~\ref{figure:stick+slip}(c)]. The shearon gets annihilated since it cannot
exists below a certain shear force needed to compensate the energy dissipation.
As a result we find that the static force $f_{\rm s}$ needed to be overcome in
order to initiate the motion is the shear force needed to create the shearon.
This becomes clearer in a stop/start experiment, where for a smoothly sliding
top plate the external drive is stopped for a certain time and reinitiated
afterwards. We find that the static friction force needed to restart the motion
manifests a stepwise behavior as a function of stopping time
\cite{Porto/Urbakh/Klafter:1999}. It vanishes only as long as the motion is
restarted within the lifetime of the shearon, giving a possible explanation of
the memory effects observed in \cite{Yoshizawa/Israelachvili:1993}.

Since in the shearon description the frictional force is determined by the wave
vector, it is suggestive to influence the friction by modifying $q$. It is
possible to change the shearon wave vector by changing external parameters such
as the stage velocity (a higher velocity results in general in a larger wave
vector). A shearon with a large wave vector, created at a high velocity, can be
maintained at lower velocities by deceleration, giving rise to a hysteretic
behavior observed in many experimental systems
\cite{Yoshizawa/McGuiggan/Israelachvili:1993+Hu/Carson/Granick:1991+%
Klein/Kumacheva:1995+Berman/Drucker/Israelachvili:1996}.

\begin{figure}[h]
\nopagebreak\vspace*{5mm}\nopagebreak
\def\epsfsize#1#2{0.619#1}
\noindent\hfill\epsfbox{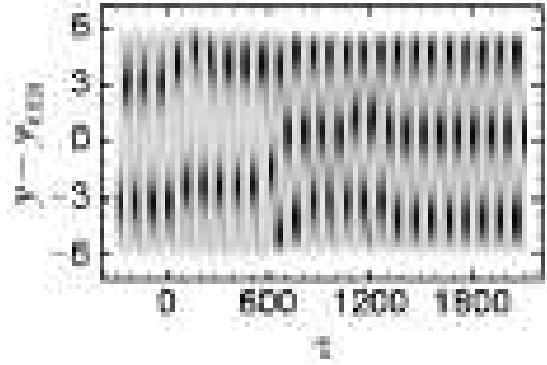}\hfill\hspace*{0mm}%
\nopagebreak\vspace*{5mm}\nopagebreak
\caption{Modifying a shearon by a defect; plotted is the particles
density $\varrho$ vs position $y-y_{\rm cms}$ and time $\tau$. The
shallow defect is reached by the first particle at $\tau \protect\cong 0$
and left by the last particle at $\tau \protect\cong 1346$. The
scattering by the defect increases the shearon wave vector, which results
in a decrease of the mean friction force from $\left< f_{\rm k} \right>
\protect\cong 0.263$ to $\left< f_{\rm k} \right>' \protect\cong 0.178$.
The model parameters are $\alpha = 1$, $\beta = 1$, $\gamma = 0.75$,
$\Delta = 0.1$, $\epsilon = 0.01$, $N = 15$ [harmonic interaction and
free boundary conditions], $v = 0.02$, and $h = 0.5$. [Please note that
the figure has a reduced quality due to figure size limitations of the
condmat database.]}
\label{figure:defect}
\end{figure}

Here, we present as an example a ``chemical'' method to manipulate the shearon
wave vector by introducing a defect. The defect is placed at the bottom plate
at an integer position $y_0$, leading to a modified bottom plate potential
$\Phi_{y_0}'(y) = \Phi_0 (1-h \{1 + \cos (2 \pi [y-y_0])\})$ for $\left|y -
y_0\right| \le 1/2$ and $\Phi_{y_0}'(y) = \Phi(y)$ otherwise. Here $h$ defines
the relative depth of the minimum at position $y_0$, with $h = 1$ being the
regular case. As an example, the resulting density pattern for $h = 1/2$ is
shown in Fig.~\ref{figure:defect}, where it can be seen that scattering by the
defect changes the shearon wave vector $q$ into a new wave vector $q' > q$. The
new shearon with wave vector $q'$ is {\it stable}, leading to a decrease of the
mean friction force from $\left< f_{\rm k} \right> \cong 0.263$ before
passing the defect to $\left< f_{\rm k} \right>' \cong 0.178$ afterwards.
Depending on the amplitude of $h$, both a decrease and an increase of the
friction force are possible, which provides a method to tune $\left< f_{\rm k}
\right>$. It was already observed in similar models that disorder can
significantly change the frictional behavior
\cite{Sokoloff:1995+Sokoloff/Tomassone:1998,%
Braiman/Hentschel/Family/Mak/Krim:1999}.

In this Letter we have introduced the concept of shearons, which are
shear-induced collective modes in the embedded system, and have demonstrated
that their properties dominate the frictional behavior of the driven system.
The results have been obtained for a certain set of parameters describing the
embedded system. However, the collective effects discussed above have a general
nature and have been found in a wide range of parameters, except for the limits
$N \to 1$, $\beta \to 0$, and/or $\Delta \to 0$, which correspond to the case
of independent particle motion. We believe that the observed collective modes
of motion are even more general and should not depend on the one-dimensionality
of our model. In particular, we expect shearons to persist in systems of higher
dimension, as long as the embedded system remains a monolayer, so that each
particle interacts with both the bottom and the top plate. The possibility to
modify shearons and hence the frictional response by {\it (i)}~(ambient)
pressure and {\it (ii)}~defects at the plates should realizable experimentally.

Financial support from the Israel Science Foundation, the German Israeli
Foundation, and DIP grants is gratefully acknowledged. M.P. gratefully
acknowledges the Alexander-von-Humboldt foundation (Feodor-Lynen program) for
financial support.

\newpage

\end{document}